# Double Blind Comparisons using Groups with Infeasible Inversion

**Abstract.** Double Blind Comparison is a new cryptographic primitive that allows a user who is in possession of a ciphertext to determine if the corresponding plaintext is identical to the plaintext for a different ciphertext held by a different user, but only if both users co-operate. Neither user knows anything about the plaintexts corresponding to either ciphertext, and neither user learns anything about the plaintexts as a result of the comparison, other than whether the two plaintexts are identical. Neither user can determine whether the plaintexts are equal without the other user's co-operation.

Double Blind Comparisons have potential application in Anonymous Credentials and the Database Aggregation Problem.

This paper shows how Double Blind Comparisons can be implemented using a Strong Associative One-Way Function (SAOWF). Proof of security is given, making an additional assumption that the SAOWF is implemented on a Group with Infeasible Inversion (GII), whose existence was postulated by Hohenberger and Molnar.

**Keywords:** Database aggregation, inference, private record linkage, privacy-preserving secure joins, Secure Multi-Party Computation, Double Blind Comparison, Groups with Infeasible Inversion, Associative One-way functions.

## 1 Double Blind Comparison

Double Blind Comparison is a new cryptographic primitive that allows a user who is in possession of a ciphertext to determine if the corresponding plaintext is identical to the plaintext for a different ciphertext held by a different user, but only if both users co-operate. Neither user knows anything about the plaintexts corresponding to either ciphertext, and neither user learns anything about the plaintexts as a result of the comparison, other than whether the two plaintexts are identical.

Double Blind Comparison differs from Secure Multi-Party Computation [33]. With Secure Multi-Party Computation, the users each know their own secret and compute some function of the combined secrets, without revealing any information about their own secrets to the other parties; whereas with Double Blind Comparison, the users do not know the secret messages they are comparing.

This is an important distinction, as it allows two users to determine whether or not their two secrets are equal. If the secret is known to one of the users, then the knowledge that the two secrets are equal immediately reveals the value of the other user's secret. This rules out Secure Multi-Party Computation as a solution.

This paper shows how Double Blind Comparisons can be implemented using a Strong Associative One-Way Function (SAOWF). Proof of security is given, making an additional assumption that the SAOWF is implemented on a Group with Infeasible Inversion (GII), whose existence was first postulated jointly by Hohenberger [17] and Molnar [22].

## 2 Preliminaries

### 2.1 Terminology

Following Homan [18], we use the phrase "2-ary function" to mean "two argument function" and the phrase "1-ary function" to mean "one-argument function."

For any 2-ary function $\sigma$, we will use prefix and infix notation interchangeably; i.e. $\sigma(x, y) \equiv x\sigma y$. Where the meaning is clear, we will omit the function symbol $\sigma$ entirely.

A function $\sigma: G \times G \to G$ is *total* if it is defined over every element in its domain; i.e. if $\forall\, x, y \in G, \sigma(x, y) \in G$. Unless explicitly stated as being partial, all 2-ary functions in this paper are total over $\Sigma^* \times \Sigma^*$.

A 2-ary function $\sigma: G \times G \to G$ is *associative* if $\forall\, x, y, z \in G, (x\sigma y)\sigma z = x\sigma(y\sigma z)$.

A 2-ary function $\sigma: G \times G \to G$ is *one-way* if, given $z \in G$, it is infeasible to find $x, y \in G$ such that $\sigma(x, y) = z$. Saxena and Soh [31] call this property *weakly one-way*.

A 2-ary function $\sigma: G \times G \to G$ is *strong* if (i) given $x, z \in G$, it is infeasible to find $y \in G$ such that $\sigma(x, y) = z$, and (ii) given $y, z \in G$, it is infeasible to find $x \in G$ such that $\sigma(x, y) = z$. Saxena and Soh [31] call this property *strongly one-way*.

An Associative One-Way Function (AOWF) is a 2-ary function $\sigma$ that is associative and one-way. A SAOWF is a 2-ary function $\sigma$ that is strong and associative.

Early papers ([28][29]) assumed that a strong function was necessarily one-way. In 2000, however, Hemaspaandra, Pasanen, and Rothe [14] showed that if $P \neq NP$, *strongness* does not imply *one-wayness,* and vice-versa: *one-wayness* does not imply *strongness*.

In this paper, we state explicitly that the function in question is Strong, Total, and Associative, but not necessarily One-way.

## 2.2 Perfect Indistinguishability

We follow the terminology of Damgard and Nielsen [9].

Consider a probabilistic polynomial time (PPT) algorithm $U$. If we run $U$ on input string $x$, the output will be a probability distribution: for every possible string $y$ there some probability that $y$ is output when $x$ was the input. We call this probability $U_x(y)$. $U_x$ is the probability distribution of $U$'s output, on input $x$.

Next, consider two probabilistic algorithms $U, V$. We run both $U$ and $V$ on the same input $x$, and we choose one of the outputs produced, which we call $y$.

Definition: Given two PPT algorithms $U, V$, we say that $U, V$ are perfectly indistinguishable, written $U \sim^p V$, if $U_x = V_x$ for every $x$.

## 3 Related work

### 3.1 Secure Multi-Party Computation

Secure multi-party computation was initially suggested by A. C. Yao in a 1982 paper [33], in which he introduced the millionaire problem: Alice and Bob are two millionaires who want to find out who is richer without either revealing to the other the precise amount of their wealth. Yao proposed a solution allowing Alice and Bob to satisfy their curiosity while respecting the constraints. In general, Secure Multi-Party Computation allows $n$ users, each of whom has a secret $S_i$, to compute some function $F(\{S_i\}_{i=1}^n)$, without allowing user $U_i$ to learn the value $S_j$ for any $j \neq i$.

Because each user $U_i$ knows his or her own secret $S_i$, one problem that cannot be solved with secure multi-party computation is whether $S_i = S_j$ for any $j \neq i$, as this would immediately reveal the value of $S_j$ to user $U_i$.

With Double Blind Comparisons, neither of the users knows the value of the secret they are comparing, and thus the question of equality can be resolved without compromising the secret.

## 3.2 Associative One-way Functions and Groups with Infeasible Inversion

Rabi and Sherman first proposed the idea of Associative One-Way Functions (AOWF) and SAOWF [28] [29]. These papers showed how AOWF and SAOWF could be used for secret key agreement and digital signatures. In [28], the question of whether such functions existed was an open problem. In [29], Rabi and Sherman published their results formally and proved that AOWFs exist if and only if $P \neq NP$, leaving the question of whether SAOWFs exist as an open problem.

Beygelzimer, Hemaspaandra, Homan and Rothe [1] presented a survey of work on AOWFs up to 1999.

Hemaspaandra and Rothe [15] proved that Strong Total Commutative Associative One-way Functions exist if and only if $P \neq NP$. Hemaspaandra and Rothe states without proof, and incorrectly, that strongness implies one-wayness. However, Hemaspaandra, Pasanen, and Rothe [14] subsequently showed that if $P \neq NP$, *strongness* does not imply *one-wayness,* and vice-versa: *one-wayness* does not imply *strongness*. Even though $\sigma(x,y)$ is *one-way*, knowledge of $x$ (resp. $y$) provides additional information that may make it easier to recover $y$ (resp. $x$); conversely, even though $\sigma(x,y)$ is *strong*, this does not preclude the possibility of finding a pair $v, w \in G$, where $v \neq x, w \neq y$, such that $\sigma(v, w) = z$.

Subsequent to this, Hemaspaandra, Rothe and Saxena [16] considered the four properties: Strong, Total, Commutative, Associative; and for each property, considered three cases: True, Not True, and "Don't Care". This resulted in $3^4 = 81$ possible cases. Hemaspaandra, Rothe and Saxena showed that if standard one-way functions exist, then 2-ary one-way functions exist for each of those 81 cases.

Homan [18] proved that:

- if standard, unambiguous, one-way functions exist, then there exist strong, total, associative, one-way functions that are O(n)-to-one;
- there exists an ($n^{O(1)}$)-to-one, strong, total AOWF if and only if P ≠ FewP, where FewP is the class of languages in NP that are accepted by machines with a polynomial-bounded number of accepting paths on each input;
- no O(1)-to-one total, associative functions exist in $\Sigma^* \times \Sigma^* \to \Sigma^*$; and
- for every nondecreasing, unbounded, total, recursive function g: N → N, there is a g(n)-to-one, total, commutative, associative, recursive function in $\Sigma^* \times \Sigma^* \to \Sigma^*$.

Hohenberger [17] and Molnar [22] studied the importance of imposing group structure on one-way functions. They jointly define a Group with Infeasible Inversion (GII) to be a standard group $G$ with the additional property that $\forall x \in G$, finding $x^{-1}$ is difficult, and show that

i) a SAOWF over a group forms a GII; and
ii) the group operator of any GII is a SAOWF.

Lemma 1 [17][22]: A SAOWF $\sigma: G \times G \to G$ over a group $G$ forms a GII.

Proof: All the standard group properties are satisfied by assumption, so it only remains to show that $\forall x \in G$, finding $x^{-1}$ is difficult.

Suppose conversely there is a PPT algorithm $A$ that, given $x \in G$, finds $x^{-1}$ with non-negligible probability. Then there exists a PPT algorithm $A'$ that, given $x, z \in G$, finds $y \in G$ such that $xy = z$, as follows: $A'$ calls $A$ with input $x$ and receives output $x^{-1}$ with non-negligible probability. $A'$ computes $y = x^{-1}z$; then $xy = x(x^{-1}z) = z$, which violates the assumption that $\sigma$ is strong.

□

Lemma 2 [17][22]: If G is a GII with group function σ: G × G → G, then σ is a SAOWF.

Proof: Associativity holds because G is a group. It only remains to show that (i) given x, z ∈ G, finding y ∈ G such that xy = z is difficult, and (ii) given y, z ∈ G, finding x ∈ G such that xy = z is difficult.

(i) Suppose conversely that there is a PPT algorithm $A$ that, given $x, z \in G$, finds $y \in G$ such that $xy = z$ with non-negligible probability. Then we construct a PPT algorithm $A'$ that, given any $x \in G$, finds $x^{-1}$ with non-negligible probability, as follows:

$A'$, given any $x \in G$, selects $r \in_R G$ and calls $A$ with input $rx, r$. With non-negligible probability, $A$ returns $y$ such that $rxy = r$. It follows that $y = x^{-1}$, violating the assumption that $G$ is a GII.

(ii) Follows similarly.

□

The question of whether GIIs exist remains open. Saxena and Soh [31] defined a SAOWF to be an abelian group $(G, *)$ with a Strong Associative group function; and noted that this definition of SAOWF is equivalent to a GII. Saxena and Soh used the term "strongly one-way" to mean "strong", and "weakly one-way" to mean "one-way", as we have defined the terms here, and proposed a candidate function, which they claim is strong, that relies on a black-box oracle to compute the function; the oracle takes as input $(x, y)$ and produces a single output $\sigma(x, y)$. Saxena and Soh referred to this as an *Oracle Strong Associative One-Way Function (O-SAOWF)*.

Zucker [34] proposed a commutative SAOWF with identity, based on knot theory (specifically braid closures), but this has not been analyzed further.

## 4 Double Blind Comparisons

### 4.1 Overview

In discussing this system, we consider the following parties: a Submitter (Alice), who initiates a comparison; a Comparer (Bob), who carries out the comparison; and a Trusted Central Authority (Ted).

Note: In this paper, we distinguish between a Submitter and a Comparer. The Submitter uses *Left-encrypted* entries, which we define to be entries of the form $\{\alpha_L a_L M, a_L\}$, while the Comparer uses *Right-encrypted* entries, which we define to be entries of the form $\{M b_R \beta_R, b_R\}$. A Double Blind comparison can only be carried out between a Submitter and a Comparer.

If $G$ is abelian, then this distinction is moot. None of the proofs of security require $G$ to be either abelian or non-abelian.

We assume throughout that Alice is the Submitter and Bob is the Comparer unless otherwise noted. The subscripts $L$ and $R$ will be omitted where it is clear from the context which is the Submitter and which is the Comparer.

### 4.2 Set-up

Let $G$ be a GII with group operator $\sigma: G \times G \to G$. In the following, we omit explicit mention of the operator $\sigma$. Alice is the Submitter, Bob is the Comparer, and Ted is a semi-trusted third party.

Let $M$ be a message known only to Ted.

Let $\alpha, \beta \in G$ be Alice's and Bob's private keys respectively. $\alpha, \beta$ are chosen randomly and kept secret by Alice and Bob respectively.

Alice chooses $l_R \in_R G$ and sends $l_R$ to Ted.

Ted chooses $l_T \in_R G$ and computes and sends $\{l_R l_T M, l_T\}$ to Alice. The use of $l_T$ guards against the possibility that Alice knows an inverse for $l_R$ and has chosen it in an attempt to recover the value $M$.

Alice chooses $l_L \in_R G$ and computes $a = l_L l_R l_T$. She stores $P_\alpha = \{\alpha l_L (l_R l_T M), a\} = \{\alpha a M, a\}$. The use of $l_L$ prevents Ted from associating $P_\alpha$ with $M$.

Similarly, Bob computes and stores $V_\beta = \{(M r_T r_L) r_R \beta, b\} = \{M b \beta, b\}$, where $b = r_T r_L r_R$ is similarly constructed between Bob and Ted.

### 4.3 Double Blind Comparison

Define $P_\alpha \approx V_\beta$, where $P_\alpha = \{(\alpha a M_a), a\}$ and $V_\beta = \{(M_b b \beta), b\}$, if $M_a = M_b$; i.e. if both $P_\alpha$ and $V_\beta$ are encryptions of the same message $M$.

To determine whether $P_\alpha \approx V_\beta$, Alice chooses a random number $r$ and sends Bob $\{r(\alpha a M_a), r \alpha a\}$.

Bob calculates:

$$B_0 = (r \alpha a M_a)(b \beta)$$
$$B_1 = (r \alpha a)(M_b b \beta)$$

Bob accepts that $P_\alpha \approx V_\beta$ iff $B_0 = B_1$.

#### 4.3.1 Submitter/Comparer

If $G$ is abelian, any Submitter certificate is also a Comparer certificate.

If $G$ is non-abelian, it is still possible for one person to play the role of either Submitter or Comparer by requesting both a Left-encrypted ciphertext and a Right-encrypted ciphertext of the same message $M$.

Unless otherwise specified, we assume that a person is either a Submitter or a Comparer.

#### 4.3.2 Comparisons Initiated by Comparer

The terminology of Submitter/Comparer implies that a request for a comparison is submitted by the Submitter and answered by the Comparer. By symmetry, the operation could be reversed – there is no reason why a Comparer could not initiate the request and the Submitter perform the comparison. However, in this paper we will assume that only Submitters initiate requests.

#### 4.3.3 Multiple Trusted Central Authorities who are not mutually trusted

It is not necessary for all the encrypted secrets to be issued by the same central authority.

For example, suppose Carol is a CIA agent who is issued a CIA Submitter certificate by Ted, while Konstantin is a KGB handler who is issued a KGB Comparer certificate by Tomasz. Nothing in the protocol precludes this; the expected result would be that, when comparing secrets, Carol and Konstantin discover only that their respective secrets are not equal.

## 5 Security

We prove the following:

1. False positives: suppose Carol is issued a token $P_\theta$ by Ted, while Viktor is issued a token $V_\varphi$ by Tomasz. Clearly, it would be undesirable to discover that $P_\theta \approx V_\varphi$. We show the possibility of Ted and Tomasz issuing tokens which happen to be equivalent to be negligible.

2. False negatives: if Alice is issued $P_\alpha$ and Bob is issued $V_\beta$ by Ted, where $P_\alpha \approx V_\beta$, we show the probability of Bob mistakenly rejecting Alice also to be negligible.
3. Unrecoverability: No one who has access to $P_\alpha$ is able to recover the secret value $M$.
4. Unlinkability (1): Ted, knowing a value $M$ and given a ciphertext $P_\alpha$, but lacking knowledge of $\alpha$, has no advantage in determining whether $M$ is the plaintext corresponding to $P_\alpha$.
5. Unlinkability (2): No one, given $P_\alpha$ and $V_\beta$, is able to determine whether $P_\alpha \approx V_\beta$ without knowledge of both $\alpha$ and $\beta$.

## 5.1 False positives

False positives: suppose Carol is issued a token $P_\theta$ by Ted, while Viktor is issued a token $V_\varphi$ by Tomasz. The probability that $P_\theta \approx V_\varphi$ is negligible.

Proof: Suppose Carol is issued a token $P_\theta = \{\theta c M, c\}$ by Ted, while Viktor is issued a token $V_\varphi = \{Nk\varphi, k\}$ by Tomasz. Suppose further that, when Carol initiates a DBC protocol exchange to Viktor, Viktor accepts that $P_\theta \approx V_\varphi$.

Then it must be the case that Carol has chosen a random element $r \in_R G$ and sent $\{r\theta c M, r\theta c\}$ to Viktor, and it must also be the case that $B_0 = (r\theta c M)(k\varphi) = (r\theta c)(Nk\varphi) = B_1$, from which we can conclude that $M = N$ (since $G$ is a group).

Thus, in order for a false positive to occur, both Ted and Tomasz must have randomly chosen the same element of $G$ for the hidden message. The possibility of this happening is $\frac{1}{|G|}$, which is negligible.

□

## 5.2 False negatives

False negatives: if Ted issues $P_\alpha$ to Alice and $V_\beta$ to Bob, where $P_\alpha \approx V_\beta$, then the probability that Bob mistakenly rejects Alice is negligible.

Proof: Suppose $P_\alpha \approx V_\beta$, where $P_\alpha = \{(\alpha a M_a), a\}$ and $V_\beta = \{(M_b b \beta), b\}$. Then by definition, $M_a = M_b$.

$\forall r \in G, B_0 = (r\alpha a M_a)(b\beta) = (r\alpha a M_b)(b\beta) = (r\alpha a)(M_b b \beta) = B_1$

In which case, Bob accepts Alice.

□

## 5.3 Unrecoverability

There are two cases of interest. In the first place, an attacker (Alice) tries to recover the value of $M$ without any knowledge of the value of $M$. In the second, an attacker (Ted) who knows all possible values of $M_i$, tries to tries to recover the value of $M$ from a particular encrypted instance.

Case A1: Given $\alpha$, $P_\alpha(M_a) = \{(\alpha a M_a), a\}$, Alice cannot find any $M'$ such that $P_\alpha(M_a) = \{(\alpha a M'), a\}$.

Proof: Follows from the fact that the function is strong.

□

Case T1: Given $M_i$, $P_\alpha(M_i) = \{(\alpha a_i M_i), a_i\}$, Ted cannot find any $\alpha'$ such that $P_\alpha(M_i) = \{(\alpha' a_i M_i), a_i\}$,

Proof: Follows from the fact that the function is strong.

□

### 5.4 Unlinkability (1)

Case T2: Ted knows two secrets, $M_0$ and $M_1$, and is given $P_\alpha(M_i) = \{(\alpha a_i M_i), a_i\}$, where $i \in \{0,1\}$, but does not know Alice's secret key $\alpha$; Ted tries to determine $i$.

Formalize this as: Given $M_0$, $M_1$, and $P_\alpha(M_i) = \{(\alpha a_i M_i), a_i\}$, where $i \in \{0,1\}$; let T be any PPT algorithm that outputs a guess for $i$. Then:

$\Pr(T(M_0, M_1, P_\alpha(M_i)) \to k \in \{0,1\}: k = i) = \Pr(T(M_0, M_1, P_\alpha(M_i)) \to k \in \{0,1\}: k = 1 - i)$

Proof: For any input $M_0, M_1, P_\alpha(M_i) = \{\alpha a M_i, a\}$, there exists $\alpha'$ such that $P_\alpha(M_i) = \{(\alpha' a M_{1-i}), a\}$

Let $\alpha' = \alpha a M_i M_{1-i}^{-1} a^{-1}$.

Then $\alpha' a M_{1-i} = (\alpha a M_i M_{1-i}^{-1} a^{-1})(a M_{1-i}) = \alpha a M_i$

Then:

Pr(T guesses correctly)

$= \Pr(T(M_0, M_1, P_\alpha(M_i)) \to k \in \{0,1\}: k = i)$

$= \Pr(T(M_0, M_1, \{\alpha a M_i \alpha a M_i, a\}) \to k \in \{0,1\}: k = i)$

$= \Pr(T(M_0, M_1, \{\alpha' a M_{1-i}, a\}) \to k \in \{0,1\}: k = i)$

$= \Pr(T \text{ guesses incorrectly})$

Since the output of T is independent of the value of $\alpha$, the probability that T guesses correctly equals the probability that T guesses incorrectly; therefore,

$\Pr(T(M_0, M_1, P_\alpha(M_i)) \to k \in \{0,1\}: k = i) = \Pr(T(M_0, M_1, P_\alpha(M_i)) \to k \in \{0,1\}: k = 1 - i)$

□

Case T3: Ted knows one secret, $M_i$, where $i \in \{0,1\}$, and is given $P_\alpha(M_0) = \{(\alpha a_0 M_0), a_0\}$, and $P_\alpha(M_1) = \{(\alpha a_1 M_1), a_1\}$; Ted tries to determine $i$.

Formalize this as: Given $P_\alpha(M_0) = \{(\alpha a_0 M_0), a_0\}, P_\alpha(M_1) = \{(\alpha a_1 M_1), a_1\}$, and $M_i$, where $i \in \{0,1\}$; let T be any PPT algorithm that outputs a guess for $i$. Then:

$\Pr(T(P_\alpha(M_0), P_\alpha(M_1), M_i) \to k \in \{0,1\}: k = i) = \Pr(T(P_\alpha(M_0), P_\alpha(M_1), M_i) \to k \in \{0,1\}: k = 1 - i)$

Proof:

WLOG suppose $i = 0$.

For any $M_0, a_0, P_\alpha(M_0) = \{\alpha a_0 M_0, a_0\}$ and $M_1, a_1, P_\alpha(M_1) = \{\alpha a_1 M_1, a_1\}$,

there exist $\alpha', M'$, such that

i) $\alpha a_0 M_0 = \alpha' a_0 M'$; and

ii) $\alpha a_1 M_1 = \alpha' a_1 M_0$

as follows:

Let $\alpha' = \alpha a_1 M_1 M_0^{-1} a_1^{-1}$. Then $\alpha' a_1 M_0 = (\alpha a_1 M_1 M_0^{-1} a_1^{-1})(a_1 M_0) = \alpha a_1 M_1$

Let $M' = a_0^{-1}(\alpha')^{-1} \alpha a_0 M_0$. Then $\alpha' a_0 M' = (\alpha' a_0)(a_0^{-1}(\alpha')^{-1} \alpha a_0 M_0) = \alpha a_0 M_0$

Let $N_0 = M', N_1 = M_0$.

Then Pr(T guesses correctly)

$= \Pr(T(P_\alpha(M_0), P_\alpha(M_1), M_0) \to k \in \{0,1\}: k = 0)$

$= \Pr(T(\{\alpha a_0 M_0, a_0\}, \{\alpha a_1 M_1, a_1\}, M_0) \to k \in \{0,1\}: k = 0)$

$= \Pr(T(\{\alpha' a_0 M', a_0\}, \{\alpha' a_1 M_0, a_1\}, M_0) \to k \in \{0,1\}: k = 0)$

$= \Pr(T(\{\alpha' a_0 N_0, a_0\}, \{\alpha' a_1 N_1, a_1\}, N_1) \to k \in \{0,1\}: k = 0)$

$= \Pr(T \text{ guesses incorrectly})$.

Since the output of T is independent of the value of $\alpha$, the probability that T guesses correctly equals the probability that T guesses incorrectly; therefore,

$\Pr(T(P_\alpha(M_0), P_\alpha(M_1), M_i) \to k \in \{0,1\}: k = i) = \Pr(T(P_\alpha(M_0), P_\alpha(M_1), M_i) \to k \in \{0,1\}: k = 1 - i)$

□

### 5.5 Unlinkability (2)

Given two encrypted values, $P_\alpha(M) = \{\alpha a M, a\}$ and $V_\beta(N) = \{Nb\beta, b\}$, no process can determine whether $M = N$ without knowledge of both $\alpha$ and $\beta$.

Formally, given any PPT algorithm T with input $P_\alpha(M_0) = \{\alpha a_0 M_0, a_0\}$, $P_\alpha(M_1) = \{\alpha a_1 M_1, a_1\}$ and $V_\beta(M_i) = \{M_i b_i \beta, b_i\}$, where $i \in \{0,1\}$, and output $k \in \{0,1\}$,

$\Pr(T(P_\alpha(M_0), P_\alpha(M_1), V_\beta(M_i)) \to k \in \{0,1\}: k = i)$

$= \Pr(T(P_\alpha(M_0), P_\alpha(M_1), V_\beta(M_i)) \to k \in \{0,1\}: k = 1 - i)$

We analyze the following attacks:

Case A2: Alice knows $P_\alpha(M_0) = \{\alpha a_0 M_0, a_0\}$, $P_\alpha(M_1) = \{\alpha a_1 M_1, a_1\}$, $V_\beta(M_i) = \{M_i b_i \beta, b_i\}$, where $i \in \{0,1\}$, and her own secret key $\alpha$. Alice succeeds if, without Bob's co-operation, she has an advantage in determining $i$.

Formalize this as: Given $\alpha, P_\alpha(M_0) = \{(\alpha a_0 M_0), a_0\}, P_\alpha(M_1) = \{(\alpha a_1 M_1), a_1\}$, and $V_\beta(M_i) = \{M_i b_i \beta, b_i\}$, where $i \in \{0,1\}$; let T be any PPT algorithm that outputs a guess for $i$. Then:

$\Pr(T(P_\alpha(M_0), P_\alpha(M_1), V_\beta(M_i), \alpha) \to k \in \{0,1\}: k = i)$

$= \Pr(T(P_\alpha(M_0), P_\alpha(M_1), V_\beta(M_i), \alpha) \to k \in \{0,1\}: k = 1 - i)$

Proof:

WLOG suppose $i = 0$.

For any $M_0, a_0, P_\alpha(M_0), M_1, a_1, P_\alpha(M_1)$, and $V_\beta(M_0)$ there exists $\beta'$ such that $M_0 b \beta = M_1 b \beta'$:

Let $\beta' = (b^{-1})(M_1^{-1})(M_0 b \beta)$

Let $N_0 = M_1, N_1 = M_0, c_0 = a_1, c_1 = a_0$

Then $P_\alpha(M_1) = P_\alpha(N_0) = \{\alpha c_0 N_0, c_0\}$, $P_\alpha(M_0) = P_\alpha(N_1) = \{\alpha c_1 N_1, c_1\}$, and $V_\beta(M_0) = V_{\beta'}(N_0) = \{N_0 b \beta', b\}$,

Since the output of T is independent of $\beta$,

Pr(T guesses correctly)

$= \Pr(T(P_\alpha(M_0), P_\alpha(M_1), V_\beta(M_0)) \to k \in \{0,1\}: k = 0)$

$= \Pr(T(P_\alpha(N_1), P_\alpha(N_0), V_{\beta'}(N_0)) \to k \in \{0,1\}: k = 0)$

= Pr(T guesses incorrectly).

□

Case A3: Alice knows $V_\beta(M_0)$, $V_\beta(M_1)$, $P_\alpha(M_i)$, where $i \in \{0,1\}$, and her own secret key $\alpha$. Alice succeeds if, without Bob's co-operation, she has an advantage in determining $i$.

Proof: Follows automatically from case B1, below.

□

Case B1: Bob knows $P_\alpha(M_0)$, $P_\alpha(M_1)$, and his own secret key $\beta$. He receives a query from Alice, $\{r\alpha a_i M_i, r\alpha a_i\}$, and determines that this corresponds to $V_\beta(M_i) = \{M_i b_i \beta, b_i\}$, where $i \in \{0,1\}$. Bob succeeds if, without Alice's co-operation, he has an advantage in determining $i$.

Formalize this as: Given $P_\alpha(M_0), P_\alpha(M_1),$ }, $V_\beta(M_i)$, where $i \in \{0,1\}$, and $\beta$; let T be any PPT algorithm that outputs a guess for $i$. Then:

$\Pr(T(P_\alpha(M_0), P_\alpha(M_1), V_\beta(M_i), \beta) \rightarrow k \in \{0,1\}: k = i)$

$= \Pr(T(P_\alpha(M_0), P_\alpha(M_1), V_\beta(M_i), \beta) \rightarrow k \in \{0,1\}: k = 1 - i)$

Proof:

WLOG suppose $i = 0$.

For any $M_0, a_0, P_\alpha(M_0), M_1, a_1, P_\alpha(M_1)$, and $V_\beta(M_0)$, there exist $\alpha', M'$ such that:

i)      $\alpha' a_0 M' = \alpha a_0 M_0$;
ii)      $\alpha' a_1 M_0 = \alpha a_1 M_1$

Let $\alpha' = (\alpha a_1 M_1) M_0^{-1} a_1^{-1}$

Let $M' = a_0^{-1} {\alpha'}^{-1} (\alpha a_0 M_0)$

Let $N_0 = M', N_1 = M_0, c_0 = a_1, c_1 = a_0$.

Then $P_\alpha(M_0) = \{\alpha a_0 M_0, a_0\} = \{\alpha' c_1 N_1, c_1\} = P_{\alpha'}(N_1)$

And $P_\alpha(M_1) = \{\alpha a_1 M_1, a_1\} = \{\alpha' c_0 N_0, c_0\} = P_{\alpha'}(N_0)$.

Since the output of T is independent of $\alpha$,

Pr(T guesses correctly)

$= \Pr(T(P_\alpha(M_0), P_\alpha(M_1), V_\beta(M_0)) \rightarrow k \in \{0,1\}: k = 0)$

$= \Pr(T(P_{\alpha'}(N_1), P_{\alpha'}(N_0), V_\beta(N_1)) \rightarrow k \in \{0,1\}: k = 0)$

= Pr(T guesses incorrectly).

□

## 6 Applications of Double Blind Comparison

This paper identifies two potential applications of Double Blind Comparison – as a possible new approach to the Database aggregation and inference problem, and as a possible new form of Anonymous Credential.

These applications will form the subject of a future paper.

## 6.1 Database aggregation and inference problem

NCSC Technical Report 005, Volume 1/5 [24] defines the inference problem as the problem "of users deducing (or inferring) higher level information based upon lower, visible data", and the aggregation problem as the problem that occurs "when classifying and protecting collections of data that have a higher security level than any of the elements that comprise the aggregate".

Denning et al. [11] identified two different types of aggregation, which White, Fisch and Pooch [32] refers to as *cardinal aggregation* and *inference aggregation*.

Cardinal aggregation occurs when an adversary collects a large number of similar records, each of which by itself is of little importance, but which by sheer volume become sensitive. An example of this used in the literature (e.g. [24]) is the CIA telephone directory, where each individual entry is of little significance, but the entire directory is considered classified.

Inference aggregation occurs when multiple databases are joined, creating virtual records with a large number of fields. While the sensitivity of the individual records may have been analyzed and found to be low, the sensitivity of the resulting virtual records has likely never been analyzed, and may be quite high.

Consider the following two databases: a military personnel database, containing two fields - a military Service Number (SN) and a job classification – and a military medical database, containing the employee SN and a medical status.

Neither of these databases appears to contain information of operational value, and both may be accorded less protection as a result. However, consider the records for the member with SN C55-111-555. The personnel database shows this person to be a CF-18 Fighter Pilot. The medical database shows this person to be 4 months pregnant.

Neither of these facts, in isolation, provides any information about the operational effectiveness of the person's military unit. When combined, however, they allow an adversary to infer that there is a CF-18 fighter pilot who is currently unavailable to fly combat missions – information that might be of operational significance.

Database inference may be made more difficult by increasing the difficulty of aggregation. For example, a government database may identify individuals by a Social Insurance Number, a bank by a customer ID number, and a business by an employee ID. Records from one database cannot then easily be linked to records from another. However, there are many cases where it is necessary to match a record from one database to its corresponding record in another.

Double Blind Comparisons allow us to distribute a large dataset among multiple databases, while retaining the ability to match and compare records across two or more of these databases. For example, an employee's personnel records can be matched to her medical profile, but only if the administrators for the personnel and medical databases co-operate. Because the records cannot be linked otherwise, the security requirements on the two databases can be made much less stringent. Even a complete compromise of both databases would not yield any useable aggregation of data.

## 6.2 Anonymous Credentials

An electronic credential is issued to a user by one organization (the issuer) that enables the user to demonstrate to a third party (the Comparer) that the user possesses some attribute. With anonymous credentials [4][5][6][7], a user is able to obtain a credential from an issuing organization (possibly using a pseudonym) and use that credential to prove possession of some attribute to multiple Comparers, using different pseudonyms, in such a way that the issuer of the credential and the Comparers would be unable to link the transactions.

Double Blind Comparisons might enable one organization (Alice) to prove to another organization (Bob) that a user (Carol) possesses some attribute or attributes (e.g. a Secret security clearance and a cryptographic public key) without revealing any more information to either Alice or Bob than is strictly necessary.

# 7 Conclusions

This paper introduces a new cryptographic primitive, Double Blind Comparisons, which allows two co-operating users, each of whom is in possession of an encrypted message, to compare those messages for equality or non-equality of the underlying plaintexts, even though neither user knows the plaintext messages. It presents a method of achieving this primitive using SAOWFs, which are known to exist if $P \neq NP$. A proof of security is provided, which requires the additional assumption that the SAOWF is over a group; the group properties are required for this proof of security, but not for the implementation of the primitive. Two potential applications, in which Double Blind Comparisons may prove useful, are identified.

## 7.1 Open problems

### 7.1.1 Splitting of Credentials

As it stands, a Submitter (Alice) can easily "split" her Submitter certificate into multiple copies and share them with her friends, as follows:

Alice's certificate is $P_\alpha = \{(\alpha a M_a), a\}$. She wants to create a new certificate to give to her friend Carol. Alice chooses two random elements $r, s \in_R G$; she sets $\gamma = r$ and $c = s\alpha a$, and computes $P_\gamma = \{rs(\alpha a M_a), s\alpha a\} = \{\gamma c M_a, c\}$. She sends Carol $\{\gamma, P_\gamma\}$.

Carol can use the newly formed pair $\{\gamma, P_\gamma\}$ directly to authenticate to any Comparer certificate $V_\beta$ where $P_\gamma \approx V_\beta$. Alternatively, she can repeat the process to come up with a new pair, $\{\delta, P_\delta\}$, in which case even Alice will no longer know Carol's secret key.

In general, the ability to split an anonymous credential is considered a highly undesirable feature. Preventing the splitting of anonymous credentials is an area of ongoing research. A number of techniques have been proposed to prevent, discourage, or limit the splitting of anonymous credentials, including PKI-assured [6][12][13][21], "All-or-nothing" [5], Biometric-based [3][19][25][26][27], and Credential Modification [2][8]. An analysis of these techniques to determine whether they can be applied to DBC-based anonymous credentials will be the subject of further research.

### 7.1.2 Unlinkability (1)

Under Unlinkability (1), we considered two cases for Ted: one in which Ted knows $M_0$ and $M_1$, and is given $P_\alpha(M_i)$, where $i \in \{0,1\}$, and another in which Ted knows $M_i$, and is given $P_\alpha(M_0), P_\alpha(M_1)$. The observant reader will have noticed that we did not address the case in which Ted knows $M_0$ and $M_1$, and is given $P_\alpha(M_i), P_\alpha(M_{1-i})$.

Indeed, the proof technique given in the text breaks down in this event and we were unable to prove the protocol secure against this attack. However, we were also unable to identify any attacks which might be used to exploit this situation, and we conjecture that the protocol is secure against this attack as well.

# References

[1]  A. Beygelzimer, L. Hemaspaandra, C. Homan and J. Rothe. One-way functions in Worst-Case Cryptography: Algebraic and Security Properties are on the House. SIGACT News 30(4), pp. 25-40, 1999

[2] M. Blanton. Online Subscriptions with Anonymous Access. ASIACCS08 (2008)
[3] G. Bleumer. Biometric yet Privacy Protecting Person Authentication. Proc. 2 IWIH, LNCS 1525, pp. 99-110. (1998)
[4] S. Brands. Rethinking Public Key Infrastructures and Digital Certificates: Building in Privacy. Ph.D Thesis. MIT Press, 2000
[5] J. Camenisch, A. Lysyanskaya. An Efficient System For Non-Transferable Anonymous Credentials With Optional Anonymity Revocation. EUROCRYPT 2001, LNCS 2045, pp. 93–118, 2001
[6] R. Canetti, M. Charikar, S. Rajagopalan, S. Ravikumar, A. Sahai and A. Tomkins. Non-transferable anonymous credentials. United States Patent 7222362. (Filing date 05/15/2000; patent publication date 05/22/2007)
[7] D. Chaum. Security Without Identification: Transaction Systems To Make Big Brother Obsolete. Communications of the ACM Volume 28 Number 70, 1985
[8] L. Chen, A. Escalante, H. Lohr, M. Manulis, and A-R. Sadeghi. A Privacy-Protecting Multi-Coupon Scheme with Stronger Protection Against Splitting. FC2007 and USEC 2007, LNCS 4886, pp. 29–44. (2007).
[9] I. Damgard and J. B. Nielsen. Commitment Schemes and Zero-Knowledge Protocols. Lectures on Data Security, LNCS 1561, pp. 63-68, 1999
[10] D. E. Denning. Cryptography and Data Security, Addison-Wesley, Reading, MA, 1982.
[11] D. E. Denning, S. G. Akl, M. Heckman, T. F. Lunt, M. Morgenstern, P. G. Neumann, and R. R. Schell. Views for Multilevel Database Security. In IEEE Transactions on Software Engineering, Vol. 13, No. 2, pp. 129-140, February 1987
[12] C. Dwork, J. Lotspiech and M. Naor. Digital signets: Self-enforcing protection of digital information. Proc 28th Annual ACM Symposium on Theory of Computing (STOC) (1996)
[13] O. Goldreich, B. Pfitzman, and R. Rivest. Self-delegation with controlled propagation – or - what if you lose your laptop. Crypto '98, LNCS 1642, pp 153-168 (1998)
[14] L. Hemaspaandra, K. Pasanen, and J. Rothe. If $P \neq NP$, then some strongly noninvertible functions are invertible. Fundamentals of Computation Theory, LNCS 2138, pp 162-171 (2001)
[15] L. Hemaspaandra and J. Rothe. Creating strong, total, commutative, associative one-way functions from any one-way function in complexity theory. Journal of Computer and System Sciences, 58(3):648–659, 1999.
[16] L. Hemaspaandra, J. Rothe, and A. Saxena. Enforcing and defying associativity, commutativity, totality, and strong noninvertibility for one-way functions in complexity theory. In ICTCS, 2005.
[17] S. Hohenberger. The Cryptographic Impact of Groups with Infeasible Inversion. Master`s Thesis, Massachusetts Institute of Technology, 2003.
[18] C. Homan. Low Ambiguity in Strong, Total, Associative, One-Way Functions. Univ. of Roch. Dept. of Computer Science, Technical Report 734. February 23, 2007
[19] R. Impagliazzo and S. Miner More. Anonymous Credentials with Biometrically-Enforced Non-Transferability. WPES 2003, ACM SIGSAC (2003)
[20] W. Lorimer. Double Blind Comparisons: A New Approach to the Database Aggregation Problem. Submitted to ACSAC 2010.
[21] A. Lysyanskaya, R. Rivest, A. Sahai and S. Wolf. Pseudonym Systems (Extended Abstract). SAC'99, LNCS 1758, pp. 184-199. (1999)
[22] D. Molnar. Homomorphic Signature Schemes. BA Thesis, Harvard College, 2003
[23] NCSC-TG-010 - A Guide to Understanding Security Modeling in Trusted Systems ("The Aqua Book"). (1992)
[24] NCSC Technical Report 005, Volume 1/5 [NCSC-TR-005-1]: Inference and Aggregation Issues In Secure Database Management Systems. May 1996. Technical report available online at http://citeseerx.ist.psu.edu/viewdoc/summary?doi=10.1.1.33.7335 (1996)
[25] S. Pape. Templateless Biometric-Enforced Non-Transferability of Anonymous Credentials. Book of Abstracts of the 2nd Weekend of Cryptography, pp. 51-55, July 2008
[26] S. Pape. Embedding Biometric Information into Anonymous Credentials. Extended Abstracts of the Second Privacy Enhancing Technologies Convention (PET-CON 2008.1), pp. 15-21, (2008)
[27] S. Pape. A Survey on Untransferable Anonymous Credentials (extended abstract). Pre-Proceedings of the Fourth IFIP / FIDIS Summer School, "The Future of Identity in the Information Society", Brno, September 2008
[28] M. Rabi and A. Sherman. Associative One-way Functions: A New Paradigm for Secret Key Agreement and Digital Signatures. Technical Report TR- CS-TR-3183/UMIACS-TR-93-124, University of Maryland College Park, 1993
[29] M. Rabi and A. Sherman. An observation on associative one-way functions in complexity theory. Information Processing Letters 64 239-244, 1997
[30] A. Saxena and B. Soh. A New Paradigm for Group Cryptosystems Using Quick Keys. 11th IEEE Intl. Conf. on Networks, 2003
[31] A. Saxena and B. Soh. A New Cryptosystem Based On Hidden Order Groups. June 26, 2006
[32] Gregory B. White, Eric A. Fisch, and Udo W. Pooch. Computer System and Network Security. CRC Press, 1996.
[33] A. C. Yao. Protocols for Secure Computations (Extended Abstract) Foundations of Computing Science (FOCS) pp. 160-164, 1982
[34] M. Zucker. Studies in Cryptological Combinatorics. PhD Thesis, City University of New York, 2005